\documentclass[pra,twocolumn, superscriptaddress, aps, showpacs, amsmath, amssymb,  nofootinbib]{revtex4}

\usepackage{graphicx}
\usepackage{dcolumn}
\usepackage{bm}
\usepackage{bbm}
\usepackage{psfrag}
\usepackage{amssymb}
\newcommand{\braket}[1]{\langle{#1}\rangle} 

\begin{document}

\title{Disordered spinor Bose-Hubbard model}
\date{\today}

\author{Mateusz \L{}\c{a}cki}
\affiliation{Instytut Fizyki imienia Mariana Smoluchowskiego, Uniwersytet Jagiello\'nski, ulica Reymonta 4, 30-059 Krak\'ow, Poland}

\author{Simone Paganelli}
\affiliation{Grup de F\'{i}sica Te\`orica: Informaci\'{o} i Fen\`omens Qu\`antics, 
Universitat Aut\`onoma de Barcelona, 08193 Bellaterra, Spain}

\author{Veronica Ahufinger}
\affiliation{ICREA-Instituci\'o Catalana de Recerca i Estudis Avan\c{c}ats, Llu\'{i}s Companys 23, 08010 Barcelona, Spain}
\affiliation{Grup d'\`{O}ptica: Departament de F\'{i}sica Universitat Aut\`onoma de Barcelona, 08193 Bellaterra, Spain}

\author{Anna Sanpera}
\affiliation{ICREA-Instituci\'o Catalana de Recerca i Estudis Avan\c{c}ats, Llu\'{i}s Companys 23, 08010 Barcelona, Spain}
\affiliation{Grup de F\'{i}sica Te\`orica: Informaci\'{o} i Fen\`omens Qu\`antics, 
Universitat Aut\`onoma de Barcelona, 08193 Bellaterra, Spain}

\author{Jakub Zakrzewski}
\affiliation{Instytut Fizyki imienia Mariana Smoluchowskiego, Uniwersytet Jagiello\'nski, ulica Reymonta 4, 30-059 Krak\'ow, Poland}
\affiliation{Mark Kac Complex Systems Research Center, Jagiellonian University,
Krak\'ow, Poland}

\begin{abstract}
We study the zero temperature phase diagram of the disordered spin-$1$ Bose-Hubbard model in 
a 2-dimensional square lattice.
To this aim,  we use a mean field Gutzwiller ansatz and a probabilistic mean field perturbation theory.
The spin interaction induces two different regimes corresponding to a ferromagnetic and antiferromagnetic order.
In the ferromagnetic case, the introduction of disorder  reproduces analogous  features of the 
disordered scalar Bose-Hubbard model, consisting in the formation of a Bose glass phase between  Mott insulator lobes.
In the antiferromagnetic regime  the  phase diagram differs more from the 
scalar case.  Disorder in the chemical potential can lead to the disappearance of Mott insulator lobes with odd 
integer filling factor and, for
sufficiently strong spin coupling, to  Bose glass  of singlets  between even filling
Mott insulator lobes. 
Disorder in the spinor coupling parameter results in the appearance of a Bose glass phase 
only between the $n$ and $n+1$ lobes for $n$ odd. Disorder in the scalar Hubbard interaction inhibits 
Mott insulator regions for occupation larger than a critical value. 
\end{abstract}

\pacs{64.60.Cn,03.75.Mn,67.85.-d}

\maketitle


\section{Introduction}\label{intro}

Spinor Bose-Hubbard (BH) models describe strongly correlated lattice systems where bosons have internal angular 
momentum whose orientation in space is not externally constrained. Bosonic interactions are 
sensitive to the spin degree of freedom leading to a rich variety of orderings in the ground state at 
zero temperature.
In atomic gases, the spin degree of freedom corresponds to the manifold of degenerate -in absence of
an external magnetic 
field- Zeeman energy states associated to a given hyperfine level $F$, i.e. \ $\{|F,m_F\rangle\}$ 
where $m_F=-F,..,F$. 
In this context, we identify the spin of the atom with the hyperfine quantum number $F$. Like in the 
scalar case, ultracold 
atomic spinor interactions can be parametrised by two-body short range (s-wave) collisions. Due to the 
rotational symmetry, 
two-body collisions between atoms  depend only on their total spin and not on its orientation. Moreover, symmetry arguments 
impose that the collisions between two {\it identical} bosons in a hyperfine spin level $F$ are restricted to total even 
spin $S=2F,2F-2,...,0$. Different properties of spinor condensates in a single trap
has been discussed \cite{Ho98,Ohmi98,Law98,Ho00}.  
The confinement of the particles in a lattice leads to an enhancement of the interactions, pushing the system to a strongly correlated regime. As it happens in the scalar BH  case 
\cite{Fisher89}, the competition between the different energy scales present in spinor BH models determines the ordering 
properties -quantum phases- of the ground state.  Modifying the energy ratio between the hopping and interactions allows 
to cross a quantum phase transition between a  spinor superfluid (SF) condensate and a Mott insulator (MI) 
state \cite{Imambekov03,Tsuchiya04,Rizzi05}.

The crucial effects of the disorder in condense matter systems were advanced in the seminal contribution of  Anderson 
\cite{Anderson58}, predicting an exponential localization of all energy eigenstates of a single particle in a periodic  
potential when additional impurities are added to it. It took several years to recognize the enormous 
consequences Anderson's result had, but nowadays it is well established that disorder, 
and specifically quenched disorder 
 (i.e. frozen during the typical time scales of the system),  is an essential ingredient in 
condensed matter systems and related  topics as conductivity, transport, high-Tc superconductivity, 
neural networks, insulating phases or quantum chaos to mention few examples (see \cite{Lewenstein07} and references therein). 
Disorder is intrinsically difficult to treat firstly because, in order to characterize the system, one should average over 
different realizations of disorder which is usually a hard task. Secondly, disordered systems often develop a complex 
landscape of low energy states making the problem of minimization to find ground states very involved. Thirdly, they 
incorporate often fractal and 
ultrametric structures, all together making the problem of simulating quantum disordered systems a very complex one. 

In recent years, it has become clear that ultracold atoms offer a new paradigm
of disordered systems, due to the fact that {\it {random}} or {\it{quasi random}} disorder can be produced in these 
systems in a {\it {controlled}} and {\it {reproducible}} way. Standard methods to achieve such controlled disorder 
are the use of speckle patterns \cite{Horak98,Boiron99} which can be added to the confining potential, or optical superlattices 
created by the simultaneous presence of optical lattices of incommensurate frequencies
\cite{Roth03,Damski03,Dinier01}. Other methods include 
using an admixture of different atomic species randomly trapped in sites distributed across the sample and acting as impurities 
\cite{Gavish05,Massignan06},
or the use  of inhomogeneous magnetic fields 
which modify randomly, close to a Feshbach resonance, the scattering length of the atoms in the sample depending 
on their spatial position \cite{Gimperlein05,Chin10}.

Strongly correlated bosons in a lattice in the presence of external random potentials were first 
considered in \cite{Fisher89} where the phase diagram in the $t-\mu$  
plane of the system, $\mu$ being the chemical potential, was worked out. The three possible ground states predicted were: 
(i) an incompressible MI
 with a gap for particle-hole excitations; (ii) a gapless Bose-glass (BG) insulator with finite compressibility
and exponentially decaying superfluid correlations in space;  
and (iii) a SF 
phase with the usual off-diagonal long range order. Previously, the onset of 
superfluidity in a random potential  in 1D was studied in \cite{Ma86}, considering hard core bosons
and using a mean field theory including quantum fluctuations, and  in \cite{Giamarchi88}, where  a 
renormalization group approach was developed to study a one-dimensional system of interacting bosons in a random potential. 
In recent years it has been shown that the question of the simultaneous presence of disorder and interactions constitutes an 
important and complex many body problem that is still far from being well understood 
(for a review see \cite{Sanchez-Palencia10}).

Here we address the effects of disorder in the strongly interacting spin-$1$ BH model
in two dimensions (2D). 
For spin-$1$ systems, the short range two body collisions lead to a spin independent effective coupling strength 
$U_0$, similar to the scalar case, plus the spinor coupling  $U_2$. With the help of a Gutzwiller ansatz, supplemented 
by a perturbative mean field approach, 
we provide the phase diagram on different regimes of the phase space determined by a 
spinor coupling and the disorder.
The Gutzwiller mean field approach is known to give reasonable results for the scalar and 
the spin-$1$ SF-MI 
transition in 2D \cite{Tsuchiya04}.  It has also been used to signal in the presence of disorder, 
a BG phase in ultracold 
scalar bosonic gases \cite{Damski03} as well as diverse glassy phases in Bose-Fermi mixtures 
\cite{Ahufinger05}. 
A Gutzwiller mean field approach yield to correct ground state for small values of the spinor 
coupling, since it neglects correlations between different sites and thus is not precise enough 
in determining 
accurately the boundaries  between distinct quantum phases. Nonetheless it provides a valuable estimate on the physics 
of the system and permits easily to include the effects of disorder going beyond the homogeneous mean field approach.

One may argue that the mean field approach could work even better in the three
 dimensional case. However, the necessarily inhomogeneous, disordered systems 
 are then much harder to treat being computationally very demanding. 
 For that reason we restrict ourselves to the 2D case only as in the earlier
 studies \cite{Damski03,Sanpera04,Ahufinger05}.

Our main results can be summarized as follows. In the non disordered case, and for $U_2>0$,  we confirm previous findings 
\cite{Tsuchiya04,Imambekov03,Kimura06} consisting in: (i) a first (second) order phase transition from 
MI to SF for even (odd) occupation numbers in the region $U_2/U_0< u_c$  (with $u_c\simeq 0.2$) 
and $t\ll U_2$, where $t$ denotes the hopping; (ii) a second order phase transition from even occupation MI lobes to SF if $U_2/U_0 > u_c$.
In the presence of disorder in the chemical potential the above effects, (i) and (ii), persist 
 together with the appearance of a  BG phase between the MI lobes. 
For  $U_2/U_0>0.5$, odd occupation MI lobes disappear while even lobes survive and the corresponding 
BG is formed only by singlets between the remaining lobes.  
Also disorder can make the  odd occupation MI lobes to disappear but the BG phase is nematic if $U_2/U_0<0.5$.
Assuming disorder in the $U_2$ coupling  
we observe that the BG phase appears only between 
every second pair of lobes and we explain such a peculiar behaviour using perturbation theory in the vanishing 
tunneling limit. On the other hand, 
disorder on the spinless term of the interaction coupling reproduces qualitatively the results found for scalar gases \cite{Gimperlein05}.

The paper is organized as follows: In section \ref{sec:model} we introduce the spin-1 BH model and shortly review the  
different phases in the homogeneous case (without disorder). In Sec. \ref{sec:atomic} 
we discuss first the exact phase diagram 
in the absence of tunneling
 to grasp  the features of the MI phase. In Sec.\ref{sec:pert} we comment the perturbative results for small 
tunneling, while in \ref{sec:Mf} and \ref{sec:gzv}  we derive a mean field 
phase diagram for finite tunneling using both, mean field perturbation 
theory (MFPT) and Gutzwiller mean field approach. In section \ref{sec:general}  
we analyze in detail the effects of disorder. Two types of 
disorder are considered here, disorder on the on-site energies,  resulting from a 
random external potential, and disorder on 
the interactions both on the scalar and  the explicit spin dependent part. We calculate the  phase diagram in the 
disordered case using both, a Gutzwiller ansatz and MFPT.  Finally, in Sec. \ref{sec:summm} we present our 
concluding remarks and open questions.

\section{Bose-Hubbard model for spin-1 bosons}\label{sec:model}

Low energy spin-1 bosons loaded in optical lattices sufficiently deep so that only the lowest energy band is 
relevant can be described by the spinor BH model.  The corresponding Hamiltonian is \cite{Imambekov03}:
\begin{eqnarray} \label{eqn:BHham}
\hat{H} & = & -t\sum_{\left\langle i,j\right\rangle ,\sigma}\left(\hat{a}_{i\sigma}^{\dagger}\hat{a}_{j\sigma}
+\hat{a}_{j\sigma}^{\dagger}\hat{a}_{i\sigma}\right)+ \frac{U_{0}}{2}\sum_{i}\hat{n}_{i}(\hat{n}_{i}-1)\nonumber\\
&+&\frac{U_{2}}{2}\sum_{i}\left(\hat{\mathbf{S}}_{i}^{2}-2\hat n_{i}\right)-\mu\sum_{i}\hat{n}_{i},
\end{eqnarray}
where $\left\langle i,j\right\rangle$ indicates that the sum is restricted to nearest neighbors in the lattice and 
$\hat{a}_{i\sigma}^{\dagger}$ ($\hat{a}_{i\sigma}$) denotes the creation (annihilation) operator of a boson in the lowest
Bloch band localized on site $i$  with spin component $\sigma=0,\pm 1$. 

The first term in (\ref{eqn:BHham}) represents the kinetic energy and describes spin symmetric hopping between 
nearest-neighbor sites with site independent tunneling amplitude $t$. The second and third term account for spin 
independent and spin dependent on site interactions, respectively. These energies at site $i$ are defined as 
$U_{0,2}=c_{0,2}\int d\vec{r} w^4(\vec{r}-\vec{r}_i)$ 
with $c_{0}=4 \pi\hbar^2(a_0+2a_2)/3m$  and $c_{2}=4 \pi \hbar^2 (a_2-a_0)/(3m)$, where $a_S$ is the s-wave scattering 
length corresponding to the channel with total spin $S$ \cite{Ho98,Ohmi98} and $w(\vec{r}-\vec{r}_i)$ is the Wannier function of the 
lowest band at site $i$.  While the second term of (\ref{eqn:BHham}) is spin independent and 
equivalent to the interaction energy for
scalar bosons, the third term represents the energy associated with spin configurations within lattice sites with
\begin{equation}
\hat{\mathbf{S}}_i=\sum_{\sigma \sigma^{\prime}=0,\pm 1} \hat{a}^{\dagger}_{\sigma i} 
\vec{F}_{\sigma \sigma^{\prime}} \hat{a}_{\sigma^{\prime} i} ,
\label{si}
\end{equation}
being the spin operator at site $i$
and $\vec{F}$ the traceless spin-1 matrices. The explicit form of the spin operator $\hat{\mathbf{S}}_{i}$ reads  
\begin{eqnarray}
\hat{S}_{z} & = & \hat{n}_{1}-\hat{n}_{-1}\nonumber \\
\hat{S}_{x} & = & \frac{1}{\sqrt{2}}\left[\left(\hat{a}_{1}^{\dagger}+\hat{a}_{-1}^{\dagger}\right)\hat{a}_{0}+H.c.\right]\nonumber\\
\hat{S}_{y} & = & \frac{i}{\sqrt{2}}\left[\left(-\hat{a}_{1}^{\dagger}+\hat{a}_{-1}^{\dagger}\right)\hat{a}_{0}-H.c.\right].
\end{eqnarray}
$\hat{\mathbf{S}}$'s components obey standard angular momentum commutation relations $[\hat{S}_{l}, \hat{S}_{j}]=i\epsilon_{ljk} \hat{S}_{k}$.
Note that the spin-interaction term favors a configuration with total magnetization zero (denoted as polar and sometimes antiferromagnetic)
for $U_2>0$ and ferromagnetic for $U_2<0$ 
\cite{Ho98,Ohmi98,Law98}.
In the grand canonical approach the total number of particles is 
controlled by the last term of (\ref{eqn:BHham}) where $\mu$ is the chemical potential and  
\begin{equation}
\hat{n}_{i}=\sum_{\sigma=0,\pm1}\hat{n}_{i,\sigma},
\end{equation}                                           
is the total number of bosons on site  $i$. 
Hamiltonian (\ref{eqn:BHham}) can be straightforwardly derived from the microscopical description of bosonic atoms, with a 
hyperfine spin $F=1$, loaded in a deep optical lattice and considering the 
two-body short range (s-wave) collisions.  More details about the derivation can be found in 
\cite{Jaksch98,Ho98,Ohmi98,Koashi00,Tsuchiya04}. 
 
Notice also that, since the orbital part of the wave function in one lattice site is the product of
 Wannier functions for all the atoms, it is symmetric under permutation of any two atoms. Therefore, the spin part of the
 wavefunction should also be symmetric due to Bose statistics. This imposes  $S_i+n_i$ to be even \cite{Ying96}, being $S_i$ and
$n_i$ the quantum numbers labelling the eigenvalues of $\hat{\mathbf{S}}_i$ and $\hat{n}_{i}$. 

As in the scalar case, the spinor BH system exhibits a quantum phase transition between superfluid and 
insulating states \cite{Imambekov03,Tsuchiya04}.  In the insulating states, fluctuations in the atom number per site are suppressed 
and virtual tunneling gives rise to effective spin exchange interactions that determine a rich phase diagram in which 
different insulating phases differ by their spin correlations. 
The appearance of spin mediated tunneling transitions in the optical lattice depends clearly on the ratio between the 
different energy scales appearing on the BH Hamiltonian (\ref{eqn:BHham}). In alkalins, the scattering lengths are such that 
spin-independent interactions $U_0$ are larger than spin-dependent ones $U_2$. In such case the SF-MI
transition depends mostly on the ratio $t/U_0$. However, inside the insulating regime, the value of 
$U_2$ plays an important role if $U_2\geq t$ where it competes with the spin-exchange interactions induced by 
small fluctuations of the particle number determining the spin structure. On the contrary, if $t\gg U_2$ tunneling 
acts similarly for all spin components and the gas will behave as a strongly correlated scalar gas. 

\subsection{The phase diagram at $t=0$.}\label{sec:atomic}
To better understand the effects of spin mediated interactions when disorder is present let us first summarize the phase 
diagram  at $t=0$ (atomic limit) 
without  disorder. In this limit, the Hamiltonian reduces to the sum of independent single-site 
Hamiltonians $\hat{H}_0=\sum_i \hat{H}_{0,i}$ with
\begin{equation}\label{eqn:hatom}
\hat{H}_{0,i}=-\mu \hat{n}_{i}+\frac{U_{0}}{2}\sum_{i}\hat{n}_{i}(\hat{n}_{i}-1)
+\frac{U_{2}}{2}\sum_{i}\left(\hat{\mathbf{S}}_{i}^{2}-2\hat{n}_{i}\right).
\end{equation}
Since  $\left[\hat{n}_i,\hat{\mathbf{S}}_{i}^{2}\right]=0$, 
the eigenstates of the single site Hamiltonian can be labeled by three quantum numbers 
$\left|S_i,m_i;n_i\right\rangle$, such that:
\begin{equation}
H^0_i \left|S_i,m_i;n_i\right\rangle=E_0 (S_i,n_i,U_0,U_2,\mu) \left|S_i,m_i;n_i\right\rangle
\label{eigen_hoi}
\end{equation}
with:
\begin{eqnarray}\label{eqn:enimp}
\label{e0}
E_0 (S_i,n_i, U_0, U_2, \mu)&=& -\mu n_i+\frac{1}{2}U_0 n_i(n_i-1)\nonumber\\
&+&\frac{1}{2}U_2\left[S_i(S_i+1)-2n_i \right] .
\end{eqnarray}

From Eq.(\ref{e0}), one can deduce the structure of the ground state of the insulator phases in the limit $t=0$. 
For antiferromagnetic interactions, $U_2>0$, the minimum energy $E^{min}_0$  is attained with minimum $S_i$, its specific 
value depending of the number of atoms per site. Thus, for even filling factor, the minimum spin is zero and the state is 
described as $\left|0_i,0_i;n_i\right\rangle$ with $n_i$ even.
This state is known as spin singlet insulator \cite{Demler02}. If the atom number per site is odd, then the minimum spin per 
site is one and the state reads $\left|1_i,m_i;n_i\right\rangle$. The chemical potential region for which each of the two 
phases are the ground states can be found easily from Eq.(\ref{e0}):
\\
\noindent (i) MI with $n$ odd and spin $1$ on each lattice site is the ground 
state if $E_0(1,n)<E_0(0,n-1)$ and 
$E_0(1,n)<E_0(0,n+1)$ i.e. when  
$(n-1)U_0<\mu<nU_0-2U_2$. This sets an upper bound on the spin coupling 
$U_2/U_0\leq 0.5$ above which the
odd lobes cease to exist \cite{Demler02}.\\  
\noindent (ii) MI with $n$ even and spin $0$ on each site are ground states if  $E_0(0,n)<E_0(1,n-1)$ and 
$E_0(0,n)<E_0(1,n+1)$  leading to $(n-1)U_0-2U_2<\mu<nU_0$ for $U_2/U_0\leq 0.5$. For higher values, odd lobes do not 
exist and the stability conditions read  $E_0(0,n)<E_0(0,n-2)$ and  $E_0(0,n)<E_0(0,n+2)$. The last two conditions set 
an $n$-dependent upper bound on the maximum value of $U_2/U_0\leq (n+1/2)$.\\
The ferromagnetic side of the diagram is easily calculated imposing
an integer number of particles and realising that the minimisation of the energy implies maximum spin value i.e. $S_i=n_i$. 

\begin{figure}[h]
 \includegraphics[height=8cm,angle=-90]{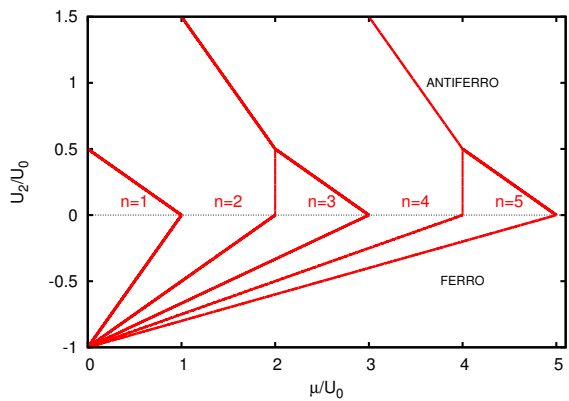}\\
 \caption{Phase diagram of the spinor $F=1$ BH model in the  limit $t=0$. Each region corresponds
to a MI phase with a different occupation number.}
 \label{fig:nodis_t0}
 \end{figure}

The exact phase diagram in the ($U_2/U_0,\mu/U_0$) plane is displayed in Fig.~\ref{fig:nodis_t0} providing the width of the 
MI lobes at $t=0$
as a function of $U_2/U_0$.
It is interesting to note that the right boundary of even lobes in the range $0<U_2/U_0<0.5$ 
does not change with $U_2$. This fact leads to a stability with respect to disorder in this parameter,
as we shall show in the Sec. \ref{sec:disU2}, corresponding to 
the absence of the BG phase between lobes with occupation $n$ and 
$n+1$ with $n$ even in the presence of disorder in $U_2$. 
In the  antiferromagnetic region, for $U_2$ large enough, odd
lobes disappear while even lobes broaden. In the ferromagnetic case the lobes shrink as $|U_2|$ increases and disappear for 
$U_2=-1$.

\subsection{Perturbative approach for small $t$}\label{sec:pert}
For small but finite tunneling $t/U_0$ and $|U_2|/U_0\ll 1$, it is possible to perform perturbation theory and derive an effective 
Hamiltonian to second order in $t/U_0$ \cite{Imambekov03}, that permits to study insulating phases.  The explicit form of the 
effective second order perturbation Hamiltonian depends on the number of bosons per site, odd or even. A mean field theory with a product
 state ansatz, applied to the effective Hamiltonian provides the following
character of insulating states for the low $t$ limit. For  $U_2<0$ the MI lobe with $n$ bosons is ferromagnetic with $S=n$. 
The situation is richer for antiferromagnetic ordering and 
depends on the dimension of the system. 

We restrict ourselves to 2D systems and revise the results from \cite{Imambekov03}. 
The MI lobes for odd $n$ are in a nematic phase  characterized by zero  expectation value of all the spin components
but broken spin symmetry $\left< S_{i z}^2\right>=0$ and 
$\left< S_{i x }^2\right>=\left< S_{i y }^2\right>=1$ (nematic phase for 3D spin systems has been studied also in \cite{Chen73,Papanicolaou88}).  The corresponding state is well described by the mean field ansatz 
$|\psi\rangle= \prod_i |S_i=1,m_i=0\rangle$. For $n$ even
(even MI lobes) and sufficiently large $U_2$, the spin-dependent term in the Hamiltonian dominates and a singlet 
configuration $|\psi\rangle= \prod_i |S_i=0,m_i=0\rangle$ is realized. However, when $t^2\propto U_0 U_2$ tunneling 
may effectively couple $S=0$ and
$S=2$ states leading again to a nematic state. As shown in \cite{Imambekov03}, a first order
transition may turn place within the MI lobe between the singlet configuration (for low $t$)
and the nematic state (at higher $t$ values) with a critical tunnelling rate fulfilling $z t_c^2=0.5U_2U_0$ for $n=2$ 
where $z$ is the number of neighbors. 
Such phase transition may take place only if $t_c$ is sufficiently small so that the MI lobe exists at
this value, otherwise a singlet MI - SF phase transition occurs first and the nematic state may not be formed.

\subsection{Standard Mean Field Perturbative Approach}\label{sec:Mf}

The MI-SF transition for spin-$1$ has also been studied using standard mean field 
perturbative approach  by Tsuchiya et al. \cite{Tsuchiya04}.  We describe these results 
in more detail since they are a starting point for our study of the effects of disorder.
Neglecting   second order fluctuations of the  bosonic annihilation and creation operators 
we obtain the condition
$(\hat{a}^\dagger_{i,\sigma}-\braket{\hat{a}_{i,\sigma}})(\hat{a}_{j,\sigma}-\braket{\hat{a}_{j,\sigma}})\simeq 0$
which allows to decouple the hopping term as
$\hat{a}^{\dagger}_{i\sigma}\hat{a}_{j\sigma}\simeq 
\psi^{*}_{i\sigma}\hat{a}_{j\sigma}+\hat{a}^{\dagger}_{i\sigma}\psi_{j\sigma}-\psi^{*}_{i\sigma}\psi_{j\sigma}$,
where we  have introduced the superfluid order parameter $\psi_{j\sigma}=\braket{\hat a_{j\sigma}}$ which, in a homogeneous 
lattice, is site-independent.
The  Hamiltonian  reduces now to a sum of local terms $\hat H_{MF}=\sum_i \hat h_i$ with 
\begin{eqnarray}\label{eqn:mfham}
\hat h& = &  -tz\sum_{\sigma}\left[\left(\psi_{\sigma}\hat{a}_{\sigma}^{\dagger}+\psi_{\sigma}^{*}\hat{a}_{\sigma}\right)
-\left|\psi_{\sigma}\right|^{2}\right]-\mu \hat{n} \nonumber\\
&+&\frac{U_{0}}{2}\hat{n}(\hat{n}-1)+\frac{U_{2}}{2}\left(\hat{\mathbf{S}}^{2}-2\hat{n}\right), 
\end{eqnarray}
where the site index $i$ has been dropped since we are considering here an homogeneous system, and $z$ denotes the 
number of nearest neighbours. 
The superfluid order parameter $\psi_\sigma$ has to be determined
by minimizing  the free energy $f=-1/\beta \log {\mathrm Tr} \left[ \exp{(-\beta \hat{h})}\right]$, where $\beta=1/K_B T$ being $K_B$ 
the Boltzmann's constant and $T$ the temperature. Here, since 
we are interested only at the zero-temperature properties, the former condition reduces to the minimization of the 
ground state energy $E_{GS}(\psi_\sigma)=\left\langle GS\right| \hat{h} \left|GS\right\rangle $ with the self-consistent 
condition
$\left\langle GS\right |\hat{a}_{\sigma} \left|GS\right\rangle=\psi_\sigma$.
For sufficiently small $t$ we can apply perturbation theory, $\hat h=\hat H_0 + \hat V(t)$, and use as a basis 
the eigenstates of $\hat H_0$ (\ref{eqn:enimp}). 
The perturbation term is given by
\begin{equation}
\hat V=-tz\sum_{\sigma}\left[\left(\psi_{\sigma}\hat{a}_{\sigma}^{\dagger}+\psi_{\sigma}^{*}\hat{a}_{\sigma}\right)
-\left|\psi_{\sigma}\right|^{2}\right].
\end{equation} 

Let us focus on the  antiferromagnetic case $U_2>0$. A tedious but straightforward calculation of the matrix 
elements of the perturbation leads to the phase boundaries between the SF phase and the MI phase in the 
$(\mu/U_0, t/U_0$) plane for a given value of the $U_2/U_0$ coupling. Notice that being 
$\hat V \propto (\hat{a}_{\sigma}^{\dagger} +\hat{a}_{\sigma})$ only even terms on the perturbation expansion 
survive.
As derived in \cite{Tsuchiya04} the ground-state energy up to
 second order is for odd occupation number given by:
\begin{eqnarray}\label{eqn:correctionodd}
& &E^{(2)}(S=1,n,t,U_0,U_2,\mu,\psi_\sigma)=\nonumber\\
&= & zt\left[1-zt\sum_{j=1,4} \alpha_j(n,U_0,U_2,\mu) \right]\sum_\sigma \left| \psi_\sigma \right|^2, 
\end{eqnarray}
and  for even occupation 
\begin{eqnarray}\label{eqn:correctioneven}
&& E^{(2)}(S=0,n,t,U_0,U_2,\mu,\psi_\sigma)=\nonumber\\
 &=&zt\left[1-\frac{zt}{3}\sum_{j=1,2} \gamma_j(n,U_0,U_2,\mu) \right]\sum_\sigma \left| \psi_\sigma \right|^2, 
\end{eqnarray}
with 
\begin{eqnarray}\label{eqn:alphas}
\alpha_1(n,U_0,U_2,\mu)&=&\frac{n+2}{3 \delta_{n-1,0;n,1} (U_0,U_2,\mu)  }, \nonumber \\
\alpha_2(n,U_0,U_2,\mu)&=&\frac{4(n-1)}{15 \delta_{n-1,2;n,1} (U_0,U_2,\mu)  }, \nonumber\\
\alpha_3(n,U_0,U_2,\mu)&=&\frac{n+1}{3 \delta_{n+1,0;n,1} (U_0,U_2,\mu)  },\nonumber \\
\alpha_4(n,U_0,U_2,\mu)&=&\frac{4(n+4)}{15\delta_{n+1,2;n,1} (U_0,U_2,\mu) }, 
\end{eqnarray}
\begin{eqnarray}\label{eqn:gammas}
\gamma_1(n,U_0,U_2,\mu)&=&\frac{n+3}{ \delta_{n+1,1;n,0} (U_0,U_2,\mu)  },\nonumber \\
\gamma_2(n,U_0,U_2,\mu)&=&\frac{n}{ \delta_{n-1,1;n,0} (U_0,U_2,\mu)  },
\end{eqnarray}
and $\delta_{l,r;n,s} (U_0,U_2,\mu)=E_0(l,r,U_0,U_2,\mu)-E_0(s,n,U_0,U_2,\mu)$. 
Minimisation of the energy for a finite order parameter (corresponding to SF) is achieved when the expressions 
on the parenthesis in (\ref{eqn:correctionodd}) and (\ref{eqn:correctioneven})  are negative. On the contrary, the 
MI phase, corresponding to zero order parameter is associated to a positive value of such expressions.
Hence, the phase boundaries between the SF and the MI in the $(\mu/U_0, t/U_0)$ 
plane, for a given value of the spin interaction $U_2$ are given by: 
\begin{equation}\label{eqn:boundaryodd}
 t_{odd}=\frac{1}{z\sum_{j=1,4} \alpha_j(n,U_0,U_2,\mu)}
\end{equation}
\begin{equation}\label{eqn:boundaryeven}
 t_{even}=\frac{3}{z\sum_{j=1,2} \gamma_j(n,U_0,U_2,\mu)}
\end{equation}
Notice also that the dimensionality of the lattice is included through the parameter $z$ which indicated the 
number of nearest neighbours.

The analysis of the ferromagnetic regime ($U_2<0$) can be done in the same way imposing the condition 
$S=n$. Since in this case all the spins are aligned, 
we can  consider only  one of the components $m=\pm S$ in the perturbative expansion.
 A straightforward calculation leads to the following explicit expression for the 
MI to SF boundary: 
\begin{equation}\label{eqn:boundaryferro}
t_{ferro}=-\frac{\left(n+n U_2-\mu \right) \left[(-1+n) \left(1+U_2\right)-\mu \right]}{z \left(1+U_2+\mu \right)}.
\end{equation}

\subsection{Variational Gutzwiller approach}\label{sec:gzv}

The variational Gutzwiller approximation is a non perturbative approach, where the 
 wave function  takes a form of a product over all $M$ sites of the lattice
\begin{equation}\label{eqn:gutwanstz}
 \left|\psi\right\rangle =\prod_{i=1}^{M}\sum_{n=0}^{n_{max}}g_{i}(n)
\sum_{S=0}^{n}f_{i}(S,n)\sum_{m=-S}^{S}h_{i}(S,m,n)\left|S,m,n\right\rangle _{i}
\end{equation}
where $g_i,\ h_i,\ f_i$ are the variational coefficients to  be determined by  minimizing the BH Hamiltonian 
(\ref{eqn:BHham})  with the above ansatz. That
implies decoupling in the tunneling term
$\langle \hat{a}_{i\sigma}^{\dagger}\hat{a}_{j\sigma}\rangle=\hat{n}_{i\sigma}\delta_{ij} + \langle \hat{a}_{i\sigma}^{\dagger}
\rangle\langle\hat{a}_{j\sigma}\rangle(1-\delta_{ij})$. 
Observe that for consistency of notation we should rather use $\psi_{i\sigma}$ instead of $\langle\hat{a}_{j\sigma}\rangle$.
The Gutzwiller variational state is a product state of on-site wave functions so it cannot reproduce intersite correlations or 
entanglement between different sites. 
 Being a generalisation of the standard mean field approximation, the Gutzwiller ansatz is expected to be exact in the limit of 
infinite dimensions. To mark the limits between the SF and MI phases
in the Gutzwiller approach, we recall that the MI phase prevails for small hopping amplitude and it is characterized by a finite 
gap in the spectrum and zero compressibility defined as $\kappa=\frac{\partial \rho}{\partial \mu}$
with 
\begin{equation}
\rho=\frac{1}{N}\left\langle \sum_j \hat{n}_j \right\rangle,
\end{equation}
and $N$  the total number of bosons. On the contrary, in the SF phase bosons are delocalized and a current flow is possible. 
This phase is characterized by a finite compressibility, gapless excitations and off-diagonal long range order accompanied by
 a non vanishing order parameter.  
Since the order parameter is not directly measurable, it is important to define experimental observable quantities 
marking the SF phase. These are typically the superfluid fraction $\rho_S$ and the condensate fraction $\rho_C$ \cite{Roth03} 
(although it has been proposed recently the measure of the compressibility directly \cite{Delande09}).
The superfluid fraction  can be evaluated 
imposing a phase gradient in the tunneling 
 corresponding to a current flow while
the condensate fraction is defined as the highest eigenvalue of the one particle density matrix \cite{Roth03,Damski03}.
In an homogeneous case the site dependence can be omitted. It is measured experimentally by means of an interference density pattern 
giving coherent peaks in the SF phase \cite{Greiner02}.
Notice that in the  mean field approach, and so in the Gutzwiller ansatz, 
the condensed fraction decouples 
and both quantities, superfluid fraction and condensed fraction are related to the average 
$\psi_\sigma = \left\langle \hat{a}_{\sigma j} \right\rangle$, 
that can be taken as 
the order parameter, its value being zero in the MI phase and finite in the SF.

\begin{figure*}
\includegraphics[height=14cm,angle=-90]{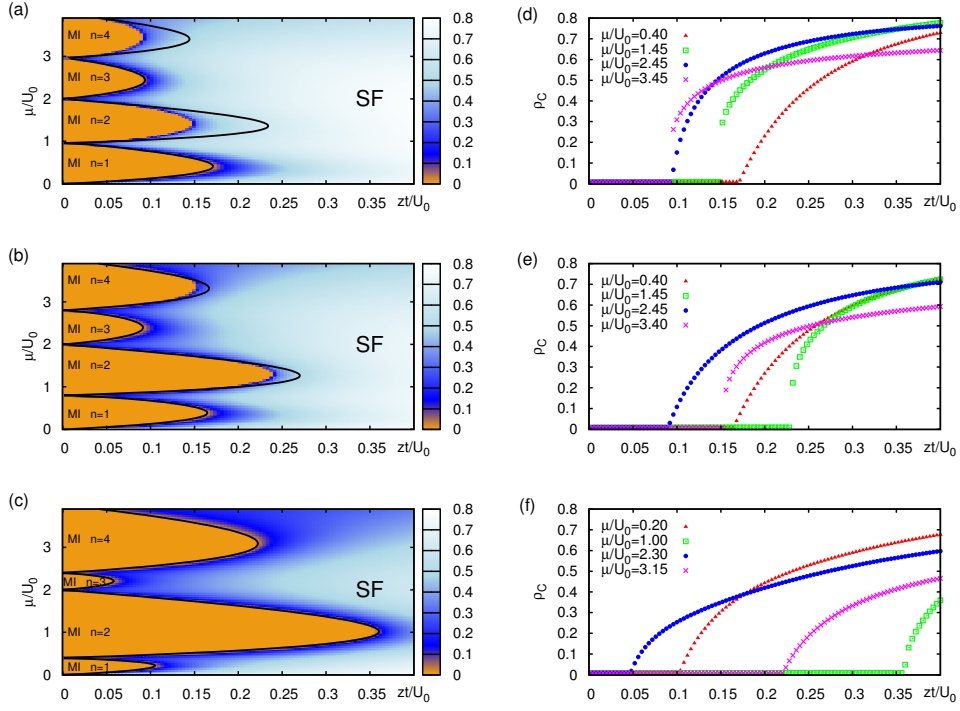}\\
\caption{(Color online). Left panels depict the condensate fraction $\rho_C$ obtained numerically by the Gutzwiller ansatz  
for, (a)$U_2/U_0=0.02$, (b)$U_2/U_0=0.1$ and (c)$U_2/U_0=0.3$, 
in the homogeneous case without disorder, where 
MI lobes correspond to vanishing  $\rho_C$ (orange areas). The lobes are compared with the boundaries obtained with 
the MFPT (solid lines). In the right panels is depicted  $\rho_C$ as a function of $zt/U_0$  for values of $\mu/U_0$ 
corresponding to the lobes' tips. 
In the transition between the MI and SF on the tip one can 
observe a first order transition for the even occupation lobes in panels (d) and (e)   (abrupt jump on the condensate 
fraction) while for lobes corresponding to odd occupation the transition is always of  the second order.}
\label{nodis}
\end{figure*}

In Figure \ref{nodis} we display the $\rho_C$  calculated both with the Gutzwiller ansatz
 and with the perturbation mean field approach (\ref{eqn:boundaryodd}-\ref{eqn:boundaryeven}) (solid line) for  
different values of the parameter  $U_2/U_0$ (left column). Observe the different behavior between odd and even lobes 
(as described in the 
previous section).  With increasing values of $U_2$,  the even lobes start to dominate while the odd lobes shrink. 
We have checked numerically that for $U_2=0.5U_0$ the odd MI lobes disappear, in agreement with the 
$t=0$ predictions of the 
previous section. For small $U_2/U_0$ ratios, there exist a discrepancy between the perturbative mean field and 
Gutzwiller predictions for the boundaries of the even lobes already reported in \cite{Kimura05}. 
This discrepancy is correlated with the character of 
MI-SF transition as visualized in the right column of Fig.~\ref{nodis}, where the condensate fraction is shown for selected 
$\mu=const$ lines corresponding to the tips of the lobes in the corresponding phase diagram. 

For $U_2/U_0 \leq 0.1$ (Fig.~\ref{nodis} first and second row)  the condensate fraction 
is continuous across the phase transition for odd lobes (corresponding to  second order phase transition) while it reveals 
a discontinuous jump, characteristic of the first order phase transition for even lobes. 
For $U_2/U_0 \geq 0.3 $ (Fig.~\ref{nodis} bottom row) only second order SF-MI transitions 
from both odd and even lobes are observed. In between these values it is not easy to  characterize the 
order of the phase transition.  An exhaustive numerical  analysis shows that the value  
where the transition passes from first to second order  is approximately $ U_2/U_0 =u_c\simeq 0.2 $.


The observation of a first order phase transition in the even lobes - where MI  is formed by singlets on each site- is not
new and has been also pointed out in the mean field analysis of \cite{Tsuchiya04, Kimura05} in 2D, as well as in 
Quantum Monte Carlo (QMC) calculations  \cite{Batrouni09} in 1D. Recall also that {\it another} first order 
transition between singlet and nematic phases has been predicted within the MI lobes \cite{Imambekov03} by using a restricted MF 
ansatz in the effective perturbative Hamiltonian.

\begin{figure}
\includegraphics[height=9cm,angle=-90]{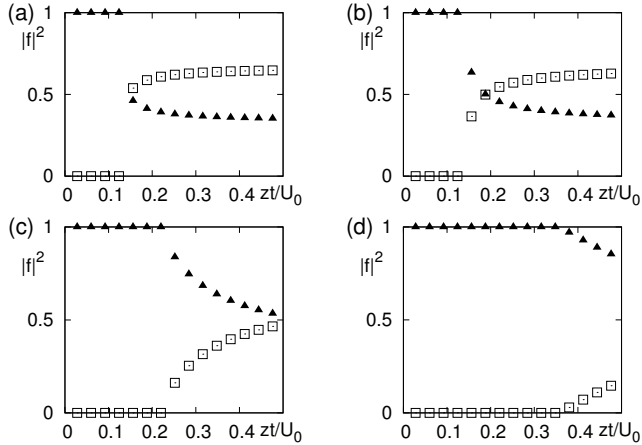}\\
\caption{Coefficients $|f(S=0,n=2)|^2$ (triangles) and $|f(S=2,n=2)|^2$ (squares) of the Gutzwiller state (\ref{eqn:gutwanstz}) 
as a function of $zt/U_0$. The value $\mu$ corresponds to the $n=2$ MI lobe's tip. The panels refer to different
values of spin interaction:  $U_2/U_0=0.01$(a), $0.02$(b), $0.1$(c) and $0.3$(d).}
\label{fig:coeffs} 
\end{figure}

Looking at the state provided by the Gutzwiller ansatz  near the even 
lobes' tips (Fig. \ref{fig:coeffs}), we observe that in the MI  only the $S=0$ component is relevant, while  
in the SF phase  the state  becomes  a linear combination of 
 $f(S=2,n) |S=2,m=0,n\rangle + f(S=0,n) |S=0,m=0,n\rangle$ with $|f(S=2,n)|^2+|f(S=0,n)|^2$  
close to $1$. A close inspection of the coefficients of the Gutzwiller ansatz (\ref{eqn:gutwanstz})  
shows that, for   $U_2/U_0<0.1$,  $f(S=2,n)$ assumes a finite value abruptly (see Fig.\ref{fig:coeffs} (a) (b) (c)).
Both states tend to contribute equally in the limit case in which $U_2= 0$ (scalar case). 
Notice that this is also the origin of the discrepancy with the MFPT result where only states with $S=0$ are taken into account in the 
energy corrections.  
Even if we find that, in the MI, the state is singlet, the second order phase transition indicates a metastability 
inside the lobe of a nematic phase. 
A rough explication of this effect can be made for the lobe corresponding to
$n=2$, noticing that
a configuration with $S=2$  starts to become favorable when the  kinetic energy becomes comparable with  
$E_0(S=2,n)-E_0(S=0,n)=3U_2$, so for  $zt_1 \simeq 6 U_2$. This value can be compared 
with the  MI tips obtained in the MFPT $zt_2=(U_0+2U_2)\left[(2n+3)-\sqrt{4 n^2+12n}\right]$. If $t_1<t_2$ 
the kinetic energy reduces the lobe with respect to the MFPT prediction and, as soon as the metastable nematic state becomes
stable, SF phase appears discontinuously. On the other hand, if $t_2<t_1$ the system becomes SF before
the appearance of the  $S=2$ contribution and the MI-SF transition is smooth.
It is easy to verify that $t_1\simeq t_2$  for  $U_2/U_0\simeq 0.145$ which is not too far from the $u_c\simeq 0.2$ mentioned above.
In contrast, for $U_2/U_0>u_c$ Gutzwiller and MFPT approaches coincide and effectively the contribution of the state $S=2$ is irrelevant
close to the tip of the lobe (see  Fig.\ref{fig:coeffs} (d)) and the transition rather than a phase crossing becomes again 
a second order phase transition.

\section{Disorder in Spinor Bose-Hubbard Model}\label{sec:general}

As discussed in the introduction, the presence of disorder in the BH model allows, apart from MI, for another
insulating phase, the BG phase. The characteristics of the phase diagram depend on the way the disorder is introduced. Here 
we will study the effect of two different kinds of disorder: disorder in $\mu$ (Sec. \ref{sec:mu}), and disorder 
in the interactions $U_2$ and $U_0$ (Sec. \ref{sec:disU2}).

Diagonal disorder can be taken into account by adding to 
the Hamiltonian (\ref{eqn:BHham}) a local term such
\begin{equation}
\hat{H}_D=\hat{H}+ \sum_i \hat{H}_{dis}(\epsilon_i)
\label{addition}
\end{equation}
where $\epsilon_i$ is a random variable defined for every site $i$  with a given probability distribution $p(\epsilon)$.
While, depending on the origin of the disorder, different  $p(\epsilon)$ may be considered
(see e.g. \cite{Delande09}).  Here we consider the simplest uniform distribution with $-\Delta \leq\epsilon_i \leq \Delta$
and the cases in which the  disorder  is equivalent to add   a random term to
one of the variables $\mu$, $U_2$ or $U_0$.

Notice that the addition of a site dependent disorder introduces inhomogeneity into the system. Thus, neither the mean field nor 
the Gutzwiller ansatz reduce to a ``single site'' effective Hamiltonian.  
Instead, the mean fields as well as Gutzwiller
wave function coefficients become explicitly site dependent. 

A Stochastic  Mean Field Theory  (SMFT),  
taking into account the  inhomogeneity  of $\psi_{i \sigma}$, has been proposed in \cite{Hofstetter2009} for the 
scalar BH. 
Here we  present a more simple MF theory, being a limiting case of  SMFT, and compare it with the 
phase diagram obtained with the Gutzwiller ansatz.  Good agreement for the MI boundaries has been found, 
 as in the non disordered case, while the BG can be seen only by the Gutzwiller ansatz.

\subsection{Probabilistic Mean Field approach} \label{sec:stoc}

A first estimation of the MI lobes in the  presence of disorder can be obtained by a MFPT,
as described in the previous section.  The generalization to the disordered case is not straightforward
since, as we mentioned, the translational invariance is broken and the order parameter  should be associated 
to a random variable  $\psi_{j \sigma}$ defined for  each site, with a certain probability distribution
$P(\psi_{j \sigma})$.  Nevertheless, since the
disorder is assumed to be homogeneous  on the lattice, we can introduce a simplified MFPT theory
taking an  average order parameter  $\bar{\psi}_{\sigma}=\int d \psi_{j \sigma}   P(\psi_{j \sigma}) \psi_{j \sigma}$.  
In doing so, we are neglecting the classical fluctuations of the order parameter induced by the disorder. 
The self-consistent condition reads  $\bar{\psi}_{\sigma}=\overline{\braket{\hat{a}_{i,\sigma}}}$, where  the 
overbar indicates an ensemble average over the lattice. It is also equivalent, due to  
the self-averaging properties of the system, to an average over the random distribution.
So the structure of the single-site mean field Hamiltonian remains the same, providing that one of the parameters changes
according to $\nu\rightarrow \nu+\epsilon_j$, where, for a diagonal disorder,  $\nu$ can be $\mu$,  $U_2$ or $U_0 $.

The minimization of the average ground-state energy which, 
up to second order corrections, reads  
\begin{eqnarray} 
\bar{E}(s,n,t,U_0,U_2,\mu,\bar{\psi}_\sigma)&=&\bar{E}_0(s,n,U_0,U_2,\mu)\\
& +&\bar{E}^{(2)}(s,n,t,U_0,U_2,\mu,\bar{\psi}_\sigma),\nonumber
\end{eqnarray}
determines if  $\bar{\psi}_{\sigma}$ is finite or zero.
Notice that, since we are neglecting the fluctuations on the order parameter,   a vanishing $\bar{\psi}_{\sigma}$ always 
corresponds to a MI phase, while BG cannot be detected, since it  has $\bar{\psi}_{\sigma}=0$ but finite fluctuations.  
A more complete analysis needs a more complex theory, such as the SMFT \cite{Hofstetter2009,Bissbort09} where 
fluctuations are taken into account and  $P(\psi_{j \sigma})$ is determined self-consistently. 

In our simplified stochastic approach the MI-SF boundary is calculated 
using the equations for the MI boundaries (\ref{eqn:boundaryodd}) and (\ref{eqn:boundaryeven}) 
but using the averaged values $\bar{\alpha}_j$ and $\bar{\gamma}_j$ instead of
$\alpha_j$ and $\gamma_j$.

\subsection{Disorder in $\mu$}\label{sec:mu}

\begin{figure*}
\includegraphics[height=14cm,  angle=-90]{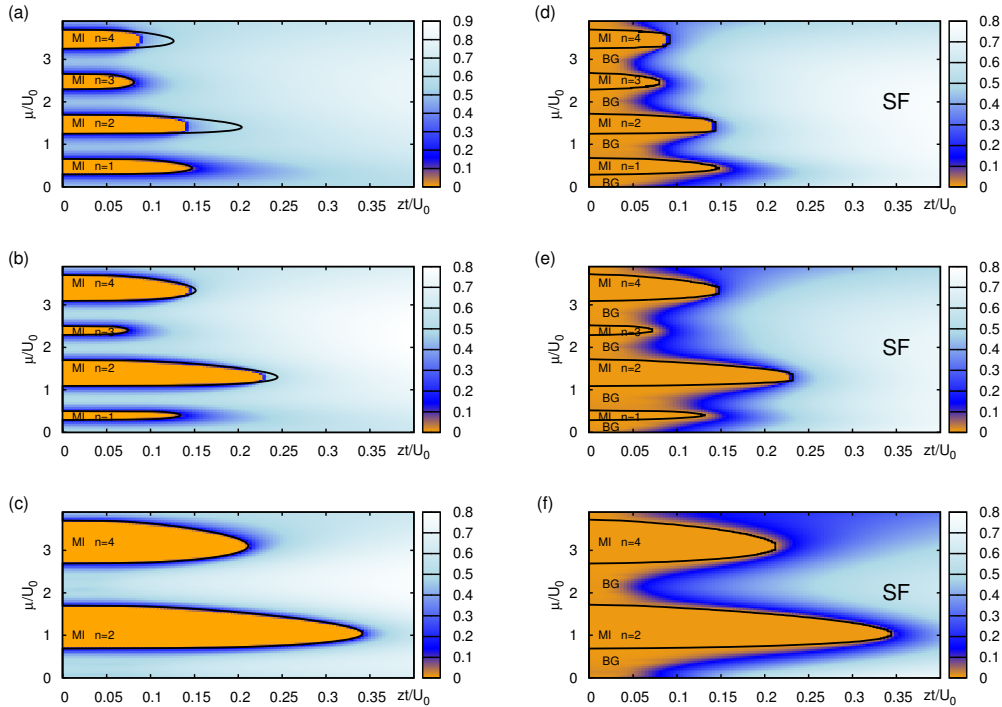}
\caption{(Color online).
Left column panels report the average density fluctuations  $\sqrt{\overline{n^2}-\overline{n}^2}$ obtained  
 from the  Gutzwiller MF approach.  MI lobes,
corresponding to vanishing fluctuations (orange areas), are compared with
 the probabilistic mean field prediction (solid lines). 
Right column panels show the corresponding condensate fraction in comparison with the Gutzwiller MI lobes (solid lines). 
The zero-condensate fraction areas (orange areas) outside the MI lobes correspond to BG phase.
For all panels, random  disorder in the chemical potential  with $\Delta=0.3U_0$ is considered. 
The different panels correspond  $U_2/U_0=0.02$ (a-d), $U_2/U_0=0.1$ (b-e) 
and $U_2/U_0=0.3$ (c-f). Observe the disappearance of the odd filling MI lobes for the largest
$U_2/U_0$ ratio in agreement with the simple estimate given in text. 
}
\label{dis}
\end{figure*}

We consider first the  disorder in the chemical potential corresponding to $\hat{H}_{dis}(\epsilon_i)=\epsilon_i \hat{n}_i$. 
To study the phase diagram, we use the Gutzwiller approach with a lattice large enough that self-averaging over the
possible disorder realizations  is already realized.
Now we have to distinguish between three phases: SF, MI and BG. As before the (disorder averaged) condensate fraction helps to find the border between SF and insulator (BG, MI) phases. 

For MI, as mentioned in the previous section, both the compressibility and the fluctuations in the average occupation number vanish within the Gutzwiller
ansatz approach. The latter simply because the MI is realized as a Fock state
with the same occupation at each site. In a BG phase the wavefunction is again a product of Fock states at each site but with different occupations (due to local action of the disorder). Thus for a BG,  fluctuations 
in the average (over sites) occupation number are significant. We have checked that we obtain numerically practically the same border between MI and BG using fluctuations in the average occupation number or by directly calculating the compressibility from its definition (see the previous section).

Let us mention that situation is so simple and unambiguous in Gutzwiller approximation only. For finite tunneling the real MI state is {\it not} a Fock state and fluctuations of on site occupation change smoothly across the MI-SF
transition (see e.g. \cite{Damski06}). In experimental situation, in addition,
atoms are held in an additional trap so the density of atoms depends on the position in the trap. Then, however, one can use directly compressibility
measurements for finding MI borders as experimentally shown for fermions
\cite{Schneider08} and also proposed for bosons \cite{Delande09}. Standard time of flight
interference patterns then allow to determine the condensate fraction.

Fig. \ref{dis} shows the results obtained for a fixed amplitude of the disorder,    
and different values of $U_2$. 
Mott Insulator lobes correspond to vanishing density fluctuations and zero compressibility as shown in the left panels (a-c).
The results obtained from MFPT are also displayed in the panels as a solid line for comparison.
As in the scalar bosonic case, disorder slightly shrinks and separates the Mott lobes and a BG phase appears between them.
The regions in the $(\mu/U_0,t/U_0)$ plane associated with the BG phase are obtained
by contrasting results obtained from the condensate fraction  ($\rho_C$) 
with the zero-density fluctuation regions (MI lobes).
The regions associated to bose glass  phase correspond to those regions where fluctuations 
are different from zero (compressible) but have vanishing condensate fraction. 
These regions  are depicted in Fig. \ref{dis} (panels d-f). In these regions the single
site superfluid parameter can be different from zero but has to vanish on average 
so to destroy the off-diagonal interference terms of the $\rho_C$.

As one can see in Fig. \ref{dis} ,
no BG appears close to the tip of a given lobe, yielding  a direct
SF-MI transition even in the presence of disorder. This is a limitation of the mean field approach.
Recently it has been claimed by means of the noninclusion 
theorem and supported by QMC calculations,   that BG always separates SF from MI phase \cite{Pollet09}. 

\begin{figure}
\includegraphics[height=8cm,angle=-90]{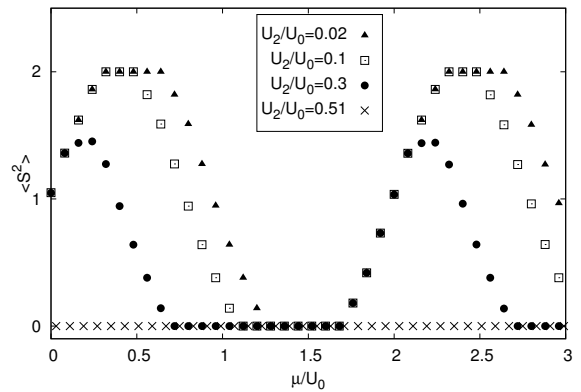}\\
\caption{Total spin average $\left\langle \hat{S}^2 \right\rangle$ as a function of $\mu/U_0$ with $zt/U_0=0.02$ for
$U_2/U_0=0.02$ (solid triangles), $U_2/U_0=0.1$ (empty squares),  $U_2/U_0=0.3$ (solid circles) and
 $U_2/U_0=0.51$ (crosses).}
\label{fig:spins} 
\end{figure}

The MI phase, in the scalar BH model, disappears completely for $\Delta>0.5U_0$. This may be 
easily understood from the fact that the maximal possible gap separating the ground state and first excited 
states is, in a homogeneous case and in $t\rightarrow 0$ limit,  equal to $U_0$. Thus disorder
spanning $[- U_0/2, U_0/2]$ interval effectively fills up the gap, producing a disordered gapless medium 
\cite{Fallani08}. The same argument may be used for odd and even lobes in the spinor case. For the 
even lobes the maximal gap  is $U_0+2 U_2$ when $U_2<0.5 U_0$ while for odd  occupation lobes
the maximal gap is $U_0-2U_2$. Thus the critical disorder for the disappearance of the odd occupation lobes is
$\Delta_o=U_0/2-U_2$.  In  Fig.~\ref{dis} the values of critical disorder are, from top to  bottom, 
$\Delta_o=0.48 U_0 $, $ 0.4 U_0$ and $ 0.2 U_0$. Then, $\Delta>\Delta_0$ only for the last case, where we see the disappearance 
of the odd occupation lobes.
It is interesting to note that  when the odd filling MI is suppressed, the BG is nematic for 
$U_2/U_0 < 0.5$  while it is formed by singlets for $U_2/U_0 > 0.5$. This fact can be seen in Fig. \ref{fig:spins} where
the averaged  $\left\langle \hat{\mathbf{S}}^{2} \right\rangle$ is plotted as a function of 
$\mu/U_0$, for $zt/U_0=0.02$ and four different values of $U_2$. 
Comparing this plot with Fig.\ref{dis}, one can see that, for $U_2/U_0 < 0.5$,  MI phases correspond to constant value of $S$, either
$\left\langle \hat{\mathbf{S}}^{2} \right\rangle=2$ (odd lobes) or $\left\langle \hat{\mathbf{S}}^{2} \right\rangle=0$ (even lobes). 
Outside this constant values 
the associated phase is BG. For $U_2/U_0=0.3$, where odd filling lobes exist in the ordered case but
are suppressed by the disorder, the BG has  $0<S<1$ ($0 < \left\langle \hat{\mathbf{S}}^{2} \right\rangle< 2 $) 
corresponding to a nematic phase 
(since $\left\langle \hat{S}^2_z \right\rangle=0$ and $\left\langle \hat{\mathbf{S}}^{2} \right\rangle \neq 0$).
On the other hand, for   $U_2/U_0>0.5$  both, the even filling MI and the BG,  have $\left\langle \hat{\mathbf{S}}^{2} \right\rangle=0$, meaning that they are formed by singlets.
So disorder can destroy insulator with odd filling, but only for  $U_2/U_0>0.5$ singlet Bose Glass is formed.

Notice that, as in the non-disordered case,
 the Gutzwiller ansatz closely coincides with the SMFA  for the boundaries of the odd occupation lobes while it 
disagrees for the even ones for sufficiently small $U_2$ (Fig.\ref{dis}). Based on the intuition
obtained from the case without disorder we may again associate the disagreement with the hidden
first order transition. Such a situation occurs for  $t<\sqrt{U_2U_0}$. For larger  $U_2$ MFPT and the 
Gutzwiller approach produce practically identical results.

\subsection{Disorder in $U_2$ and $U_0 $}\label{sec:disU2}

One could imagine that disorder in the on-site interactions $U_2$ and $U_0 $ can be experimentally realized, 
in principle, using optical Feshbach resonances \cite{Fedichev96,Bohn97,Theis04,Chin10} (the application of magnetic field in a standard Feshbach resonance technique would additionally modify the system due to e.g. Zeeman level splitting). However the optical Feshbach resonance introduces losses dues to spontaneous emission from the intermediate state \cite{Bohn97,Theis04} so it is not clear at all whether the timescale
for losses would allow for realizing the ground state of the system.  Very recently, however, another microwave-Feshbach resonance technique has been suggested \cite{Papoular10}. This method uses resonant microwave driving between ground state sublevels to tune the scattering length. Since excited states are not involved in this method no additional losses due to spontaneous emission are expected. Although this approach is up to now a theoretical proposal,
it seems to be a promising candidate for tuning the interactions
in a stable way without the application of the magnetic field.

A small local fluctuation in the laser tuning (assuming optical scheme with the reservation discussed above) or the microwave tuning (in the method of \cite{Papoular10}) $\delta \omega$  introduces
fluctuations in $U_0$ and $U_2$ so, in principle, disorder should be considered in both parameters.
Since the variations $\delta a_S$ of the two scattering lengths are correlated variables (both being
function of $\delta \omega$), we could manage 
to compensate them so to have an almost vanishing sum or difference.
If, for instance, we put the system between  the two Feshbach resonances, a small detuning will increase one scattering length and
decrease the other one. So, if the condition $\delta a_0 (\delta \omega) +2 \delta a_2 (\delta \omega)\simeq 0 $ holds,  
only disorder in $U_2$
can be considered,   on the contrary, if $\delta a_2 (\delta \omega) - \delta a_0 (\delta \omega)\simeq 0 $
 we can consider only disorder in $U_0$.

Let us start considering  disorder in $U_2$, so assuming that for each site
$U_2^i=U_2+\epsilon_i$  where $\epsilon_i$  takes a random value  in the interval $ [-\Delta,\Delta]$.
Throughout this section we assume that $\Delta<|U_2|$ so  to  consider all $U_2^i$ of the same sign, 
negative or positive, for the ferromagnetic or the antiferromagnetic cases, respectively.

\begin{figure*}
\includegraphics[height=14cm,angle=-90]{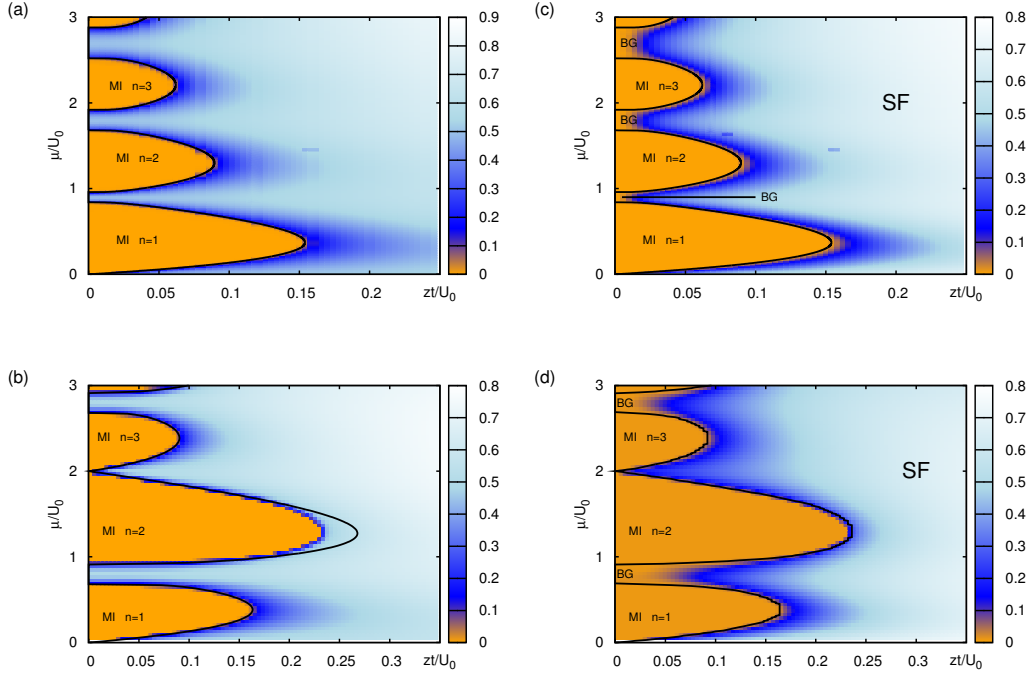}\\
\caption{(Color online). Density fluctuations (left panels) and $\rho_C$ (right panels) for $U_2=\pm0.1 U_0$ and disorder in $U_2$ $\Delta/U_0=0.06$.  
MI lobes compared with the MF results  (solid lines). 
Vanishing $\rho_C$ outside the MI lobes   (solid lines), corresponds to the BG phase.
Panels (a) and (c) correspond to the ferromagnetic case  $U_2=-0.1 U_0$ (a-c). The case $U_2=0.1 U_0$ is reported in panels (b) and (d). 
}
\label{fig:disu2}
\end{figure*}

Figure~\ref{fig:disu2} shows the effect of disorder for  $U_2/U_0=\pm 0.1$ and $\Delta/U_0=0.06$.
In the ferromagnetic case (plots (a) and (c)), disorder in $U_2$ has the same effect
as the disorder in $\mu$. Similar to the scalar case, MI lobes are shrunk and BG phases
appear between them.
In contrast new features emerge in the antiferromagnetic case (plots (b) and (d)), where  BG is formed only between 
lobes corresponding to $n$ and $n+1$
occupations with $n$-odd. No effect of disorder is visible between  $n$ and $n+1$ MI lobes for $n$-even. 
A very simple explanation of that behavior may be obtained from Fig.~\ref{fig:nodis_t0}. Note 
that for $U_2\in[0,U_0/2]$ the border separating $n$ and $n+1$
occupations for $n$-even in the $\mu - U_2$ plot is vertical (in $t=0$ limit). Thus changes (e.g. fluctuations) 
in $U_2$ do not modify the chemical potential at which the density changes. 
For $n$ odd in this range of $U_2$ 
the border is tilted, thus fluctuations in $U_2$ for a fixed $\mu$ change the density value which is favored for 
the ground state. Then, depending on the particular value
of $U_2$ at a given site the density for the ground state changes. Interestingly, this picture,
established for $t=0$, seems to hold also for finite $t$ as no BG is observed between 
the odd and even lobes. Further inspection of  Fig.~\ref{fig:nodis_t0} reveals that 
in other possible ranges of $U_2$ the lines separating different densities are always tilted -
indicating possibility of BG creation between the MI lobes. Incidentally, we can also interpret the same figure 
assuming fixed $U_2$ and fluctuating $\mu$ as  the case discussed earlier in this paper.
There are no horizontal lines in Fig.~ \ref{fig:nodis_t0}  thus
all density borders are vulnerable to fluctuations in $\mu$. This is again consistent with the
observation that for disorder in $\mu$ BG appears between all lobes.

\begin{figure*}
\includegraphics[height=14cm,angle=-90]{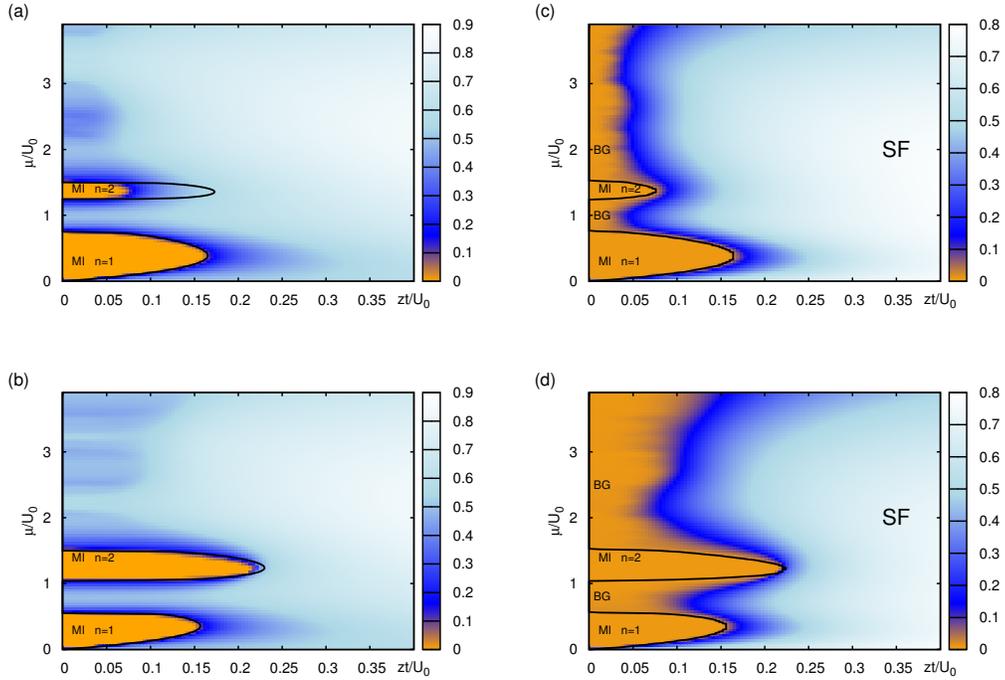}\\
\caption{(Color online). 
Density fluctuations (left panels) and $\rho_C$ (right panels) for  disorder in $U_0$ with $\Delta/U_0=0.25$ .  
MI lobes compared with the MF results  (solid lines). 
Vanishing $\rho_C$ outside the MI lobes   (solid lines), corresponds to the BG phase.
Panels (a) and (c) correspond to   $U_2/U_0=0.0$, (b) and (d)  to $U_2/U_0=0.1$. }
\label{fig:disu0}
\end{figure*}

Finally, in Fig.~\ref{fig:disu0} we show the result obtained for the disorder in $U_0$. As in the previous case 
we take  $U_0^i=U_0+\epsilon_i$  with  $\epsilon_i\in [-\Delta,\Delta]$.  The plots report 
the case with zero and finite value of $U_2$ and  $\Delta /U_0=0.25$.
As explained in \cite{Gimperlein05}, in the $U_2=0$ case,  lobes with occupation $n>(1+\Delta/U_0)/(2 \Delta/U_0)$ 
disappear while the first one remains always stable.  In our analysis  we recover this behavior even for finite $U_2$.   In both cases
lobes separate and BG appears in between. 

\section{Summary-Open questions}\label{sec:summm}

We have analyzed the effects  of disorder in the spin-1 BH model in which
the spin interaction induces two different regimes, corresponding to a ferromagnetic and antiferromagnetic order,
focusing mainly on the antiferromagnetic case, where the  phase diagram differs more from the 
scalar case.
We have considered here both, disorder introduced to the chemical potential 
(corresponding to an offset of energies at different sites) as well as disorder in the atom-atom interactions.
 As for the scalar bosons, we have observed the appearance of a compressible insulator - the BG phase - 
its character depending on the $U_2/U_0$ ratio. For small $U_2$ when Mott states with an odd number  of atoms per site (also termed  nematic
 since they have the mean
value of all components of the spin equal zero, but a non vanishing singlet projection)exist in the absence of disorder, 
 we expect the BG to be also nematic.
 For large $U_2$ however, when odd MI lobes do not exist already in the absence of disorder, we find a BG of singlets, 
a novel phase peculiar to bosons with spin.
 
Interestingly enough, in the presence of disorder on spinor coupling $U_2$, the system shows 
robustness against BG creation which does not emerge between $n$ and $n+1$ MI lobes for $n$- even. 
This is traced back to the insensitivity of the MI borders to changes in $U_2$
in the $\mu/U_0 - U_2/U_0$ plane observed in the vanishing tunneling limit.
  
This work is only the first step towards understanding disorder on lattice spinorial bosons. 
For 1D systems density matrix renormalization group (DMRG) or its variants may be used to go beyond the mean field;
 work in this direction is in progress. For 2D, similar studies may be undertaken within QMC.  

Finally, we remark  the suitability of these systems  for spin-glass studies. 
Notice that if one induces disorder in the $U_2$ coupling not preserving the ferromagnetic and anti-ferromagnetic 
character of the two-body interactions, i.e.  if  $\Delta > |U_2|$, 
a situation resembling frustration will
appear in this model with antiferro and ferro sites randomly distributed along the lattice. 
Last but not least, a more realistic model of fluctuations in the interactions should be considered taking
the details of optical Feshbach resonance into account.

\section*{Acknowledgments}

We thank  M. Lewenstein and K. Sacha for useful discussions.  
Support from Polish Government (Foundation for Polish Science), European Community, Spanish  Government (FIS2008:01236;02425, Quoit-Consolider Ingenio 2010 (CDS2006-00019))
and Catalan Government (SGR2009:00347;00343) is acknowledged.
J. Z. acknowledges hospitality from ICFO and
partial support from the advanced ERC-grant QUAGATUA. 
M.\L{}. acknowledges support from  
Jagiellonian University International Ph.D Studies in Physics of Complex 
Systems (Agreement No. MPD/2009/6).
S.P. is supported by the Spanish Ministry of Science
and Innovation through the program Juan de la Cierva.



\bibliography{bibl-bosehubbard}

\end{document}